# Reconstruction of multiple non-line-of-sight objects using back projection based on ellipsoid mode decomposition

**CHENFEI JIN,\* JIAHENG XIE, SIQI ZHANG, ZIJING ZHANG, AND YUAN ZHAO**

*Department of Physics, Harbin Institute of Technology, Harbin 150001, China*
*\*jinchenfei@hit.edu.cn*

**Abstract:** Non-line-of-sight imaging has attracted more attentions for its wide applications. Even though ultrasensitive cameras/detectors with high time-resolution are available, current back-projection methods are still powerless to acquire a satisfying reconstruction of multiple hidden objects due to severe aliasing artifacts. Here, a novel back-projection method is developed to reconstruct multiple hidden objects. Our method considers decomposing all the ellipsoids in a confidence map into several "clusters" belonging to different objects (namely "ellipsoid mode decomposition"), and then reconstructing the objects individually from their ellipsoid modes by filtering and thresholding, respectively. Importantly, the simulated and experimental results demonstrate that this method can effectively eliminate the impacts of aliasing artifacts and exhibits potential advantages in separating, locating and recovering multiple hidden objects, which might be a good base for reconstructing complex non-line-of-sight scenes.

**References and links**

1. E. Repasi, P. Lutzmann, O. Steinvall, M. Elmqvist, B. Göhler, and G. Anstett, "Advanced short-wavelength infrared range-gated imaging for ground applications in monostatic and bistatic configurations," Appl. Opt. **48**(31), 5956-5969(2009).
2. O. Steinvall, M. Elmqvist, and H. Larsson, "See around the corner using active imaging, "Proc. SPIE, **8186**, 818605 (2011).
3. A.Velten, T. Willwacher, O. Gupta, A. Veeraraghavan, M. G. Bawendi, and R. Raskar, "Recovering Three-dimensional Shape Around a Corner using Ultrafast Time-of-Flight Imaging," Nat. Commun. **3**, 745(2012).
4. V. Molebny, O. Steinvall, "Multi-dimensional laser radars," Proc. SPIE, **9080**, 908002(2014).
5. O. Katz, E. Small and Y. Silberberg, "Looking around corners and through thin turbid layers in real time with scattered incoherent light," Nat. Photonics, **6**, 549-553 (2012).
6. O. Katz, P. Heidmann, M. Fink, and S. Gigan, "Non-invasive single-shot imaging through scattering layers and around corners via speckle correlations," Nat. Photonics, **8**,784-790(2014).
7. X. Xu, X. Xie, H. He, H. Zhang, J. Zhou, A. Thendiyammal and A. P. Mosk. "Imaging objects through scattering layers and around corners by retrieval of the scattered point spread function," Opt. Express **25**(26), 32829-32840(2017).
8. F. Xu, G. Shulkind, C. Thrampoulidis, J. H. Shapiro. A. Torralba, F.N.C. Wong, and G.W. Wornell. "Revealing hidden scenes by photon-efficient occlusion-based opportunistic active imaging," Opt. Express **26**(8), 9945-9962(2018).
9. O. Gupta, T. Willwacher, A. Velten, A. Veeraraghavan, and R. Raskar, "Reconstruction of hidden 3D shapes using diffuse reflections," Opt. Express **20**(17), 19096-19108(2012).
10. M. Laurenzis and A. Velten, "Non-line-of-sight active imaging of scattered photons," Proc. SPIE, **8897**, 889706(2013).
11. M. Laurenzis and A. Velten, "Nonline-of-sight laser gated viewing of scattered photons," Opt. Eng. **53**(2), 023102 (2014).
12. F. Heide, M. B. Hullin, J. Gregson and W. Heidrich, "Low-budget Transient Imaging using Photonic Mixer Devices," ACM Transactions on Graphics, **32**(4), 45(2013).

13. F. Heide, L. Xiao, W. Heidrich, and M. B. Hullin, "Diffuse Mirrors: 3D Reconstruction from Diffuse Indirect Illumination Using Inexpensive Time-of-Flight Sensors, " in *Proceedings of IEEE Conference on Computer Vision and Pattern Recognition* (IEEE, 2014), pp. 3222-3229.
14. C. Jin, Z. Song, S. Zhang, J. Zhai, and Y. Zhao, "Recovering three-dimensional shape through a small hole using three laser scatterings," Opt. Lett. **40**(1), 52-55(2015).
15. M. Buttafava, J. Zeman, A. Tosi, K. Eliceiri, and A. Velten, "Non-line-of-sight imaging using a time-gated single photon avalanche diode, " Opt. Express **23**(16), 20997-21011(2015).
16. M. Laurenzis, J. Klein, E. Bacher, and N. Metzger, "Multiple-return single-photon counting of light in flight and sensing of non-line-of-sight objects at shortwave infrared wavelengths," Opt.Lett. **40**(20), 4815-4818(2015).
17. G. Gariepy, F. Tonolini, R. Henderson, J. Leach, and D. Faccio, "Detection and tracking of moving objects hidden from view," Nat. Photonics, **10**, 23-27(2016).
18. M. B. Hullin, "Computational Imaging of Light in Flight," Proc. SPIE, **9273**, 9273-40 (2014).
19. A. Kadambi, H. Zhao, B. Shi, and R. Raskar, "Occluded Imaging with Time of Flight Sensors," ACM Transactions on Graphics, **35**(2), 1-12(2016).
20. J. Klein, C. Peters, J. Martín, M. Laurenzis, and M. B. Hullin, "Tracking objects outside the line of sight using 2D intensity images, " Sci. Rep. **6**, 32491 (2016).
21. S. Chan, R. E. Warburton, G. Gariepy, J. Leach, and D. Faccio, "Non-line-of-sight tracking of people at long range," Opt. Express **25**(9), 10109-10117(2017).
22. V. Arellano, D. Gutierrez, and A. Jarabo, "Fast back-projection for non-line of sight reconstruction,"Opt. Express **25**(10), 11574-11583(2017).
23. X. Pan, E. Y. Sidky, and M. Vannier, "Why do commercial CT scanners still employ traditional, filtered back-projection for image reconstruction?," Inverse Probl . **25**(12), 123009(2009).
24. A. Kak and M. Slaney, "*Principles of Computerized Tomographic Imaging*" (IEEE, 1999), Chap. 5.
25. R. A. Brooks, G. H. Weiss, and A. J. Talbert, "A new approach to interpolation in computed tomography," J. Comput. Assist. Tomo. **2**, 577-585(1978).
26. M. Laurenzis and A. Velten, "Feature selection and back-projection algorithms for nonline-of-sight laser-gated viewing," J. Electron. Imaging **23**(6), 063003 (2014).
27. M. Laurenzis and A. Velten, "Investigation of frame-to-frame back projection and feature selection algorithms for non line of sight laser gated viewing," Proc. SPIE **9250**, 92500J (2014).
28. M. Laurenzis, A. Velten, and J. Klein, "Dual-mode optical sensing: three-dimensional imaging and seeing around a corner," Opt. Eng. **56**(3), 031202 (2017).
29. M. Laurenzis, F. Christnacher, and A. Velten "Study of a dual mode SWIR active imaging system for direct imaging and non-line of sight vision," Proc. SPIE, **9465**, 946509 (2015).
30. M. Laurenzis, F. Christnacher, J. Klein, M. B. Hullin, and A. Velten "Study of single photon counting for non-line-of-sight vision, " Proc. SPIE, **9492**, 94920K (2015).
31. A. Sroka, S. Chan, R. Warburton, G. Gariepy, R. Henderson, J. Leach, D. Faccio, and S. T. Lee, "Time-resolved non-sequential ray-tracing modelling of non-line-of-sight picosecond pulse LIDAR, " Proc. SPIE, **9822**, 98220L(2016).
32. K. C. Tam, and V. Perez-Mendez, "Tomographical imaging with limited-angle input," J. Opt. Soc. Am. **71**(5), 582-592(1981).

## 1. Introduction

Laser imaging techniques have been rapidly developed for many civilian and military applications, such as navigation, terrain visualization, obstacle avoidance, weapon guidance, and target recognition. In recent years, "seeing around a corner" is becoming a new capability for laser imaging which uses reflections from mirror [1,2]or glossy surfaces [3] to extend the viewing into non-line-of-sight (NLoS) conditions[4]. Hitherto, most of NLoS imaging methods require measuring and analyzing the flight time of the scattering photons traveled beyond the line of sight vision except for several methods limited in special conditions such as wavefront shaping by spatial light modulators [5], autocorrelation of the speckle pattern scattered from a diffuse wall [6], retrieval of the scattered point spread function with the aid of a reference object [7], anti-pinhole imaging using an occluder as a lens[8]. Hence NLoS imaging systems are strictly dependent on the response time of optical sensors, and then ultrasensitive cameras/detectors with high time-resolution are adopted subsequently from steak camera [9], intensified Charge-Coupled Device (ICCD) [10,11], photonic mixer devices (PMDs) [12,13] to single-photon avalanche diode (SPAD) /array [14-17].

Despite all above hardware available, an enhancement of the imaging qualities of NLoS objects is still a bottleneck because that reconstruction of the objects is mathematically considered as an ill-posed inverse problem. Approaches including convex optimization and back projection have been mainly developed in recent years. The former has advantage in good reconstruction quality. Yet, this method heavily requires the priors and a large size of the projection matrix resulting in insolubility of this inverse problem [13,18-20]. In addition, this method is actually a bulk process since it requires all the projections available before the inverse problem is solved. The latter is popularly applied for NLoS imaging because of distinct superiorities including free of assumption about the hidden scene geometry [9-11], real-time capability [17, 21, 22] as well as relatively robust to noise and erroneous data [15]. Therefore, the back projection has been a promising method for the reconstruction of the hidden objects.

The back-projection methods employed in NLoS imaging field are analogous to the those used in computational tomography(CT) field [23], and has similar problems with those of CT imaging, i.e. the issue of aliasing artifacts [24]. The so-called artifacts, generally refers to those recovered voxels around the objects with non-zero intensity but do not exist in real world. The artifacts can be seen as a by-product of back projection due to data under-sampling and the limitation of projection views [25]. The artifacts can greatly degrade the quality of the reconstruction with blurry boundaries, especially for multiple NLoS objects, the overlapping among the artifacts of different objects leads to a series of difficulties to separate objects, locate position and recover shape. In this case, current back-projection methods are powerless to acquire satisfying results [9,15,26,27].

In this paper, we present a novel method to reconstruct multiple NLoS objects using back projection based on ellipsoid mode decomposition (EMD). The basic idea of our method focuses on decomposing all the ellipsoids in a confidence map into several clusters of ellipsoids belonging to different objects, respectively. Then, each object and its artifacts can be subsequently extracted from the confidence map. As a result, each object can be individually reconstructed by a procedure of filtering and thresholding. The rest of the paper is organized as following: first, we introduce the principle of general back-projection method and theoretically develop our method. Second, we compare the simulated results of reconstructing multiple objects by using general method and our method. We also experimentally demonstrate the feasibility of our method on reconstructing multiple objects. Third, potential reasons for the comparative simulated and experimental results are further analyzed in view of artifacts.

## 2. General method

If taking a light path shown in Fig. 1 as an example, the NLoS imaging process is described as follows. A pulse laser is directed towards an image screen to form a light source point $S$. The light pulse falling at a point $P$ on the objects is diffused and returns back to the image screen. For a given image point $I_i$, a detector can receive some reflected photons by directing its narrow field of view (FOV) to the image point, and a distance geometry can be formed as described by Equ. (1).

$$\|\vec{r}_1\| + \|\vec{r}_2\| + \|\vec{r}_3\| + \|\vec{r}_4\| = ct \quad (1)$$

Where $r_2$ and $r_3$ are the distances of point $P$ toward the source point $S$ and the imaging point $I_i$, respectively. $r_1$ is the distance between the source point and the laser, and then $r_4$ is the distance between the image point and the detector. Both $r_1$ and $r_4$ are independent of the objects and can be directly measured in advance. The time intervals $t$ between the laser's synchronous trigger and the received photon are measured and counted repeatedly to form a histogram $N(S, I_i, t)$ of photon counts versus time. For a reconstruction of the objects, a large amount of time histograms should be obtained for different pairs of the source point and image points by scanning the FOV of the detector throughout the image screen.

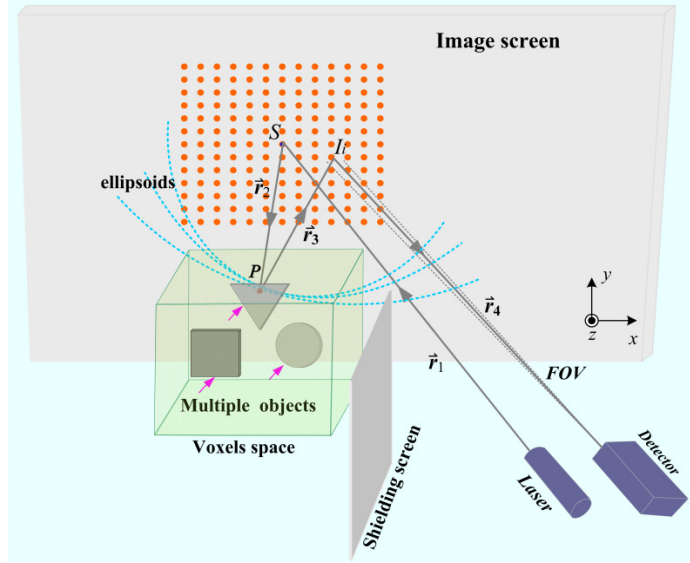

Fig. 1. The principle of NLoS imaging.

The reconstruction of the objects is based on above acquired time histograms. A back projection process is performed by projecting the photon counts in each time bin of time histogram into a voxelized Cartesian space according to the equation $V_{ij}(x,y,z)\big|_{ct_j - r_1 - r_4 - r_2 - r_3 = 0} = N(S, I_i, t_j)$. Each projection $V_{ij}(x,y,z)$ corresponds to an ellipsoid with focal points at $S$ and $I_i$, and the time bins of all time histograms are projected into the voxels space. That is to say, A mass of overlapping ellipsoids comprises a confidence map $V(x,y,z)$, in which intensity of each voxel represents the possibility of the objects occurring in this voxel. In the confidence map, every "object" consists of a certain number of intersecting ellipsoids, which is defined as an ellipsoid mode of object.

An ellipsoid mode can be divided into an object and its artifacts. The former refers to those voxels mapped to an actual object while the latter represents the rest of voxels but not existing in real world. The artifacts can be seen as a by-product of the back projection. The surface edges of the object will become blurred due to the existence of the artifacts surrounding it. Thus, a procedure of filtering and thresholding is performed for the sake of eliminating the artifacts around the object. Generally, a Laplacian filter is firstly used to enhance surface edges by computing a second derivative of the confidence map, as given in Eq. (2). Then, a thresholding algorithm written as Eq. (3) is further applied to remove artifacts and produce a 3D reconstruction of the object.

$$V_f(x,y,z) = \nabla^2 V(x,y,z) \tag{2}$$

$$V_f(x,y,z)_{> \beta \cdot \max(V_f)} = constant \tag{3}$$

The above reconstruction method is feasible to reconstruct single NLoS object but becomes inefficient to reconstruct multiple NLoS objects for serious interferences among multiple ellipsoid modes. Until now, within our knowledge, the methods suitable to multiple objects reconstruction are still lacking and will be quite crucial to NLoS imaging.

### 3. Our method

## 3.1 Back projection based on EMD

As mentioned above, the characteristic of an ellipsoid mode is the coexistence of an object and its artifacts. In case of multiple objects, the artifacts will become complex and overlapping due to the objects' different properties such as reflectivity, locations and shapes, so general method is ineffective on eliminating the artifacts and reconstructing the multiple objects. Here, we design a kind of decomposing criteria, based on which we can decompose all the ellipsoids in a confidence map into several ellipsoid modes belonging to different objects. Then, each object can be individually reconstructed from corresponding ellipsoid mode, respectively.

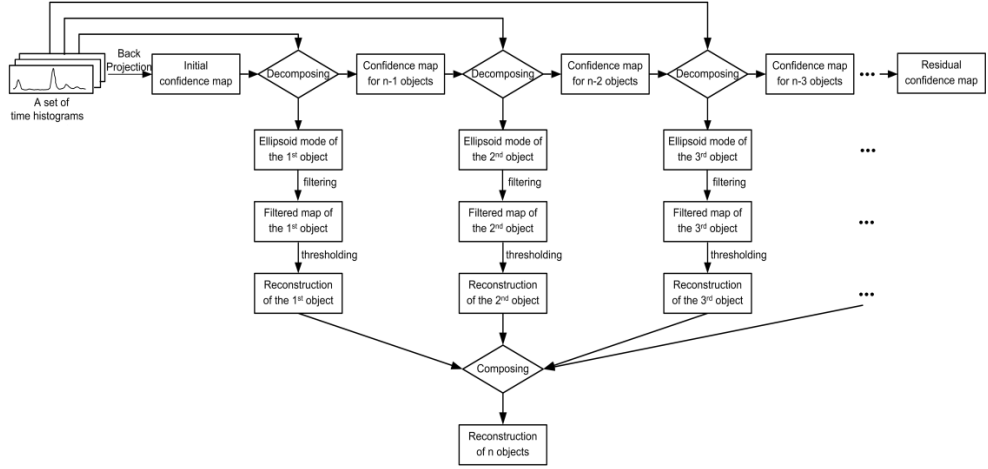

Fig. 2 Schematic diagram of back projection based on EMD

The principle of back projection based on EMD is depicted in Fig. 2. Firstly, the time histograms acquired by a NLoS system are projected into a voxelized Cartesian space to form an initial confidence map. Then decomposing is operated to extract the ellipsoid mode of the selected object from the initial confidence map and the rest is a residual confidence map as a new confidence map for next decomposing operation, and so on, the ellipsoid modes of the objects will be extracted successively by same decomposing operation until all objects have been extracted with leaving only the projections from background and noise. By multiple decomposing, the initial confidence map is divided into multiple ellipsoid modes. Next, each object can be independently reconstructed from its ellipsoid mode after filtering and thresholding. Finally, all reconstructed objects are composed into a whole reconstruction of the multiple objects. It is emphasized that decomposing operation is crucial to ensure the reconstruction quality of the multiple objects.

## 3.2 Decomposing criteria

Because the ellipsoid modes of the different objects will overlap with no obvious boundary among each other in the confidence map, it is very different to separate them using conventional image segmentation method. Our decomposing operation is based on three considerations. Firstly, it involves how the voxels are clustered into the most probable objects, respectively. Secondly, it depends on which one among the objects should be selected preferentially in this decomposing operation. Lastly, it is concerned how the ellipsoid mode of the selected object is extracted from the confidence map. Details of decomposing criteria are listed as follows.

(1) *Clustering of voxels in confidence map*

Since the objects are most likely to appear in the voxels with the maximum intensity in the confidence map, our criteria prefers to take a local maximal voxel as the central position of certain object. Considering the shape and volume of the object, those voxels around the local maximal voxel have more probability to be part of this object, estimation of which is based on two conditions. One is that the distance between an evaluated voxel and the local maximal voxel should be within certain range (spatial window), the other is the intensity difference between these two voxels should be within certain range (intensity window). Only those voxels satisfying above two conditions are clustered into the same object centering on the local maximal voxel.

(2) *Selection of preferentially extracted object*

After clustering of the voxels in the confidence map is finished, it is important to select an appropriate object that will be extracted from the confidence map in this decomposing operation. Our criteria prefer to extract "strong object" which has both high intensity and big volume. This way has two advantages. On the one hand, the "strong object" is more apt to keep real location and shape under the influence of other "weak objects", so it is easy to drawn "strong object" exactly from confidence map. On the other hand, only after the "strong object" is extracted, the "weak objects" obstructed by the "strong object" have more probability to reveal themselves in residual confidence map so that one of the appeared "weak objects" can be found and extracted in next decomposing operation.

(3) *Extraction of ellipsoid mode of selected object*

As one object is considered to be extracted, an issue about how to effectively extract the ellipsoid mode of the selected object from the confidence map becomes more important. our criteria is that only the ellipsoids passing through the voxels of the selected object are projected again to form an ellipsoid mode of the selected object, and then this ellipsoid mode is subtracted from the confidence map to leave an residual confidence map. The benefits of extracting an ellipsoid mode include two aspects: One is that the shape of the selected object can be individually reconstructed only from its ellipsoid mode that is not affected by the ellipsoid modes of remaining objects. The other is that the ellipsoid modes of remaining objects can be separated in next decomposing without the interference of this extracted ellipsoid mode.

### 3.3 procedure of decomposing

The decomposing procedure is performed by using Matlab software. List of the decomposing procedure is briefly provided in Table 1. Before decomposing, the space window ($h_s$) and the intensity window ($h_c$) are initialized.

$$\begin{aligned} |x - x_o| &\leq h_s. \\ |y - y_o| &\leq h_s. \\ |z - z_o| &\leq h_s. \end{aligned} \tag{4}$$

$$|V(x_o, y_o, z_o) - V(x, y, z)| / V(x_o, y_o, z_o) \leq h_c. \tag{5}$$

The decomposing procedure actually has a little dependence on the values of $h_s$ and $h_c$, thus the selection of $h_s$ and $h_c$ is not quite strict. In our method, $h_s$ is selected slightly larger than the size of the greatest object in the scene and $h_c$ is set as a value of about 0.4.

**Table 1 List of Decomposing Procedure**

| |
|---|
| 1  for a given confidence map $V$, initialize $h_s$ and $h_c$. |
| 2 for given any space voxel $(x_o,y_o,z_o)$, if $V(x_o,y_o,z_o)$ is the maximum value within the space window satisfying Equ.(4), find all voxels satisfying Equ. (5) within this space window. Add up the intensity of all these voxels into a sum, save the sum and the positions of these voxels. |
| 3 repeat above step 2 until all space voxels within $V$ are tested and get a list of sum value corresponding to many clusters of the positions of the voxels. |
| 4 find the maximum in the list of sum value, and then find a cluster of positions of the voxels corresponding to the maximum. |
| 5 perform back projection again but only keep the ellipsoids passing the cluster of positions of the voxels determined in step 4 to form an ellipsoid mode, and then subtract the ellipsoid mode from $V$, the rest are used as a new confidence map for next decomposing procedure. |

## 4. Results and discussion

### 4.1 Simulated results

The simulated experiment as is depicted in Fig. 1 has been performed. We assume that the image screen parallel to x-y plane is located at z=0 and positions of the laser and the detector are known in advance. The objects with complex shapes can be generated in 3D Max software and the model files are imported to MATLAB to define the positions of the objects' points. In our simulation, the round, triangle and square plate with reflectivity of 0.1, 0.3, and 1.0 are located at different distance from the image screen, respectively. The simulation process is composed of forward transmission and inverse reconstruction. In the forward transmission, the number of the photons received by the detector is obtained from the reflections of all objects points, and specific value is given by the radar equation similar to those given by the references[28-30]. Flight time of each photon is given by a ray tracing method similar to that given by the reference [31]. For a fixed source point and a set of imaging points, a set of time histograms (i.e. TCSPC data) can be acquired by the forward transmission process. There are 4024 time bins with the width of 10ps. In order to simulate the time response of the NLoS imaging system due to pulse width of the laser and time jitter of the detector, we broaden the time histograms using a Guass Kernel function with full width of half maximum (FWHM) of 50ps. In the inverse reconstruction, both general method and our method are performed with the time histograms acquired during the forward transmission process. The reconstruction region is limited in a cube of 2m×2m×1m divided into 100×100×50 voxels.

Typical results of the objects using general method are shown in Fig. 3. From Fig. 3(a), at low threshold value of 0.2, the round plate and the triangle plate can be reconstructed but with serious shape distortions, while the square plate fails to emerge obviously from its ambient artifacts. Although the square plate can be further recovered by increasing threshold value to 0.25, the round plate and the triangle plate in Fig. 3(b) are greatly distorted under this circumstance. As threshold value is increased to 0.3, quite small part of the triangle plate remains and the round plate disappears completely, while the square plate acquires an acceptable shape shown in Fig. 3(c). Typically, with threshold value up to 0.5, the square plate takes on a good shape but both the round plate and the triangle plate disappear completely shown in Fig. 3(d). From these changes, it is indicated that general method is powerless to reconstruct multiple NLoS objects with different properties such as reflectivity, locations and shapes.

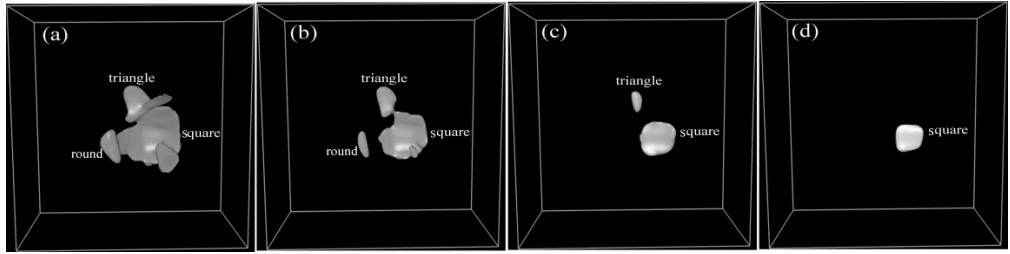

Fig. 3 (a)-(d) Simulated results using general method to reconstruct multiple NLoS objects with various threshold values such as 0.2, 0.25, 0.3 and 0.5, respectively.

The simulated results of our method are depicted in Fig. 4. It is seen that the objects are well separated and reconstructed at correct locations respectively. Especially, three objects are recovered without obvious shape distortions in contrast with the results using the general method. Compared with original objects, slight distortions are also observed and possible reasons for this phenomenon can be mainly attributed to the missing cone problem [9] which is also known to traditional CT field [32].

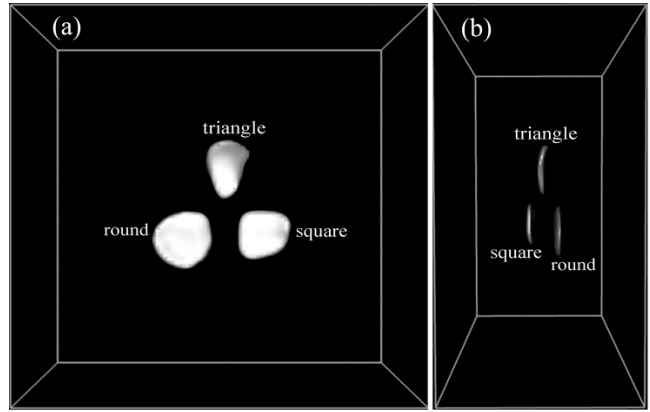

Fig. 4 Simulated results using our method to reconstruct multiple NLoS objects with (a) a view from the front and (b) a view from the side.

### 4.2 Experimental results

For the experimental proof of the theory, we constructed experimental setups as shown in Fig. 5(a). The light source is a fiber Laser emitting 90 fs light pulses at a wavelength of 1550 nm. It operates at a repetition rate of 100MHz with pulse energy of 1 nJ. By an emitting collimating lens, the fiber laser is collimated to a narrow Gaussian laser beam with 2mm diameter and 2 mrad divergence angle. The detector is a free-running InGaAs/InP SPAD with a fiber pigtail for optical input. It can provide the detection efficiency up to 25% with a time jitter of about 300 ps. A receiving collimating lens is coupled with the SPAD by a fiber pigtail, and the FOV of the detector is limited to 2 mrad. The lens is mounted on a rotating platform with adjustable elevation and azimuth angles.

In Fig. 5(b), an image screen is covered by one white paper with 5cm×5cm grids, on which the location marked by red point is selected as a source point and the locations marked by 256 blue points as image points. A shielding screen and the floor are covered by a black light-absorbing cloth for removing the scatterings from non-interesting things. A Time-correlated single-photon counting (TCSPC) unit is used to produce time histograms. Each time histogram has 1024 time bins with a width of 165 ps. The time histograms were collected with the room lights off using 90 s exposure time (high signal-to-noise ratio) to avoid the noise effect on the reconstruction quality.

The objects used in the experiment include two white rectangle plates with a size of 60 cm × 30 cm and one 30 cm × 30 cm white square plate, as shown in Fig. 5(c). All of the objects were located outside the direct field of view of the laser and the detector. The objects were placed at 0.2m, 0.35m and 0.65m distances from the image screen, respectively. The angles between the objects' plane and axes of $(x,y,z)$ were $(30°, 0°, 60°)$ for the object 1, $(0°, 30°, 60°)$ for the object 2 and $(15°, 0°, 75°)$ for the object 3, respectively.

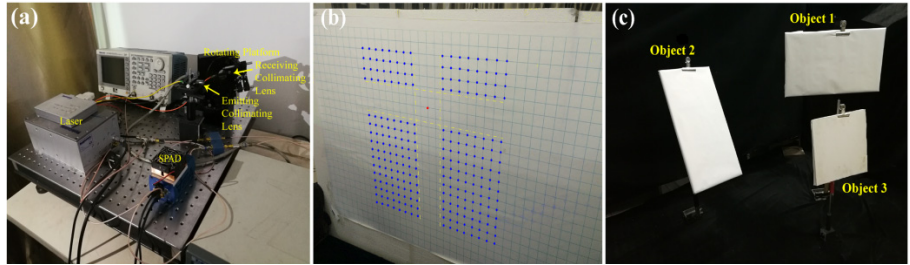

Fig. 5 Details of our experiment including (a) photograph of the experimental setups, (b) the source point and image points on the image screen, and (c) photograph of the objects.

The reconstruction results of the objects using both general method and our method are compared in Fig. 6. The reconstruction volume is limited within a box of 2m×2m×1m. Typical results using general method are shown in Fig. 6 (a) and Fig.6 (c). it is indicated that the object 2 takes on an acceptable shape, while the object 1 is concealed deep in its ambient artifacts and the object 3 almost disappears. The sizes, locations and orientations of the reconstructed objects have obvious deviations from those of real objects. In contrast, if using our method, the objects can be clearly separated and reconstructed well. As shown in Fig. 6 (b) and Fig. (d), each object takes on a good shape with size, location and orientation identical well to those of real object.

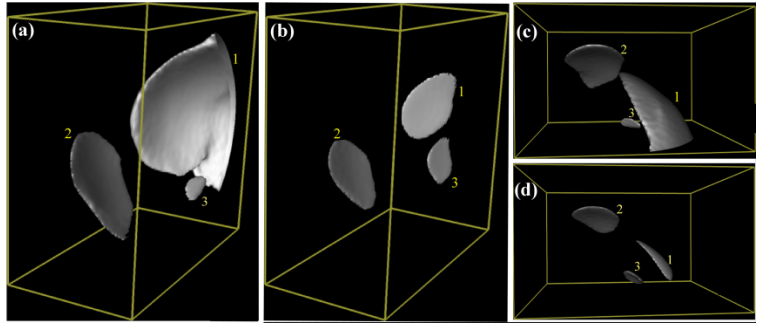

Fig. 6. Reconstruction of the objects from a side view (a) using general method and (b) using our method. Reconstruction of the objects from a top view (c) using general method and (d) using our method.

### 4.3 Analysis of simulatied results

We consider comparing the simulated results of general method with those of our method by analysis of the initial confidence map and different objects' ellipsoid modes. For general method, typical 2D slices coincided with different objects' planes in the initial confidence map are shown in Fig. 7(a)-(c). Among the objects, the square plate takes on a good shape with much higher intensity than that of ambient artifacts in Fig.7(c). In contrast, the round and the triangle plates ("weak objects") in Fig. 7(a) and Fig. 7(b) exhibit serious shape distortions and low contrasts to ambient artifacts that mainly come from the square plate. In this case, it is hard to separate the objects from aliasing artifacts even with filtering and thresholding as shown in Fig. 3.

In our method, the initial confidence map is decomposed into three ellipsoid modes belonging to different objects, respectively. The ellipsoid mode of the square plate is firstly extracted, and typical 2D slices of this ellipsoid mode coincided with three objects' planes are shown in Fig. 7(d)-(f). Fig.7 (f) indicates that the square plate can be distinguished clearly from surrounding artifacts. It is also observed that some artifacts of the square plate indeed extend to the locations of the round plate and the triangle plate marked in dotted lines in Fig. 7(d) and Fig. 7(e), which means that these artifacts are successfully integrated into the square plate's ellipsoid mode that is fully extracted from the initial confidence map.

The ellipsoid mode of the triangle plate is secondly extracted. Typical 2D slices of this ellipsoid mode at three objects' locations are depicted in Fig. 7(g)-(i). Based on the slice image shown in Fig. 7(h), the triangle plate exhibits better shape and higher contrast to ambient artifacts compared with the result shown in Fig.7(b). It is also seen that there is a blank at the location of the square plate marked in dotted lines in Fig.7 (i), which gives a proof that the ellipsoid mode of the square plate was removed so completely during last decomposing that it has little effect on current decomposing. It is also observed that small proportions of artifacts actually extend to the location of the round plate marked in dotted lines in Fig. 7(g). Similarly, it means these artifacts are successfully incorporated into the triangle plate's ellipsoid mode that is fully extracted from the initial confidence map.

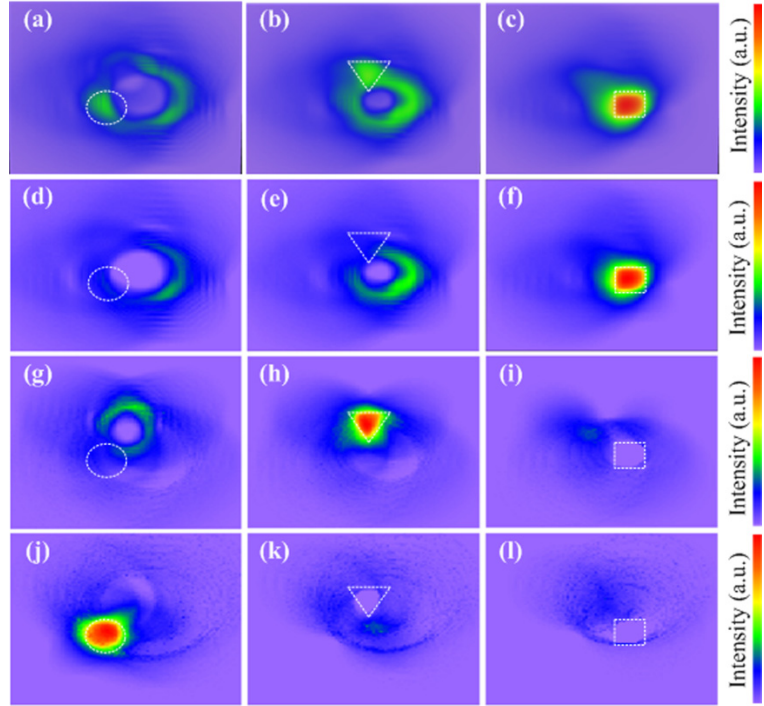

Fig. 7 (a)-(c) Typical slices of the initial confidence map at the locations of three objects, respectively. (d)-(f) Typical slices of the square plate's ellipsoid mode at the locations of the objects, respectively. (g)-(h) Typical slices of the triangle plate's ellipsoid mode at the locations of the objects, respectively. (j)-(l) Typical slices of the round plate's ellipsoid mode at the  locations of the objects, respectively.

The ellipsoid mode of the round plate is finally extracted. Typical 2D slices of this ellipsoid mode at the objects' locations are shown in Fig. 7(j)-(l). Fig.7 (j) demonstrates that the round plate take on a much better shape and much higher contrast to ambient artifacts compared with the result shown in Fig. 7(a). It is also seen that there are blanks at the locations of the triangle plate and the square plate marked in dotted lines in Fig.7 (k) and Fig.7 (l), which are also proofs that the ellipsoid modes of the triangle plate and the square plate were eliminated from the initial confidence map completely before decomposing the ellipsoid mode of the round plate. Ultimately, each object can be independently reconstructed from their own ellipsoid modes by filtering and thresholding, respectively, and then all of the recovered objects make up a reconstruction of multiple NLoS objects as shown in Fig. 4.

### 4.4 Analysis of experimental results

We also compare experimental results of general method with those of our method by analysis of the initial confidence map and different objects' ellipsoid modes. In case of using general method, the ellipsoid modes of three objects observed in Fig.8 (a) are mixed up together. Then it is quite obvious that the object 1 is the "strongest" and the object 3 is the "weakest" without clear boundary between the ellipsoid modes of these two objects. It is challenging to reconstruct all the objects even with filtering and thresholding as is shown in Fig.6 (a) and Fig.6 (c). Our method well overcomes this limitation. As are demonstrated in Fig. 8(b)-8(d), the initial confidence map is decomposed into three ellipsoid modes belonging to different objects, respectively. Especially, the ellipsoid mode of the "weakest" object 3 are separated perfectly from the ellipsoid mode of the "strongest" object 1. Each ellipsoid mode contains enough information of the objects which can be reconstructed individually by a procedure of filtering and thresholding. Finally, the experimental results shown in Fig.6 (b) and Fig.6 (d) can be obtained by combining the recovered objects with right sizes, locations and orientations together.

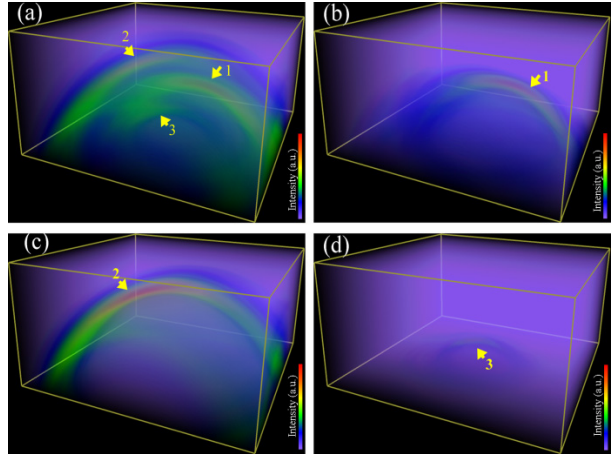

Fig. 8 (a) the initial confidence map of three objects, (b) the ellipsoid mode of the object 1, (c) the ellipsoid mode of the object 2, (d) the ellipsoid mode of the object 3.

## 5. Conclusions

For imaging of multiple NLoS objects, plenty of artifacts coexist with the objects in confidence map. In this case, current back-projection method is unsatisfying and almost hard to obtain desirable reconstruction quality. We present a novel method for reconstructing multiple hidden objects. In our method, the ellipsoid modes of different objects can be successively extracted from the initial confidence map on the basis of decomposing criteria including clustering of voxels in confidence map, a selection of preferentially extracted object and a specific extraction of ellipsoid mode. Then, each object can be independently reconstructed from its ellipsoid modes by a procedure of filtering and thresholding. Results of both simulation and experiment have demonstrated the advantages of our method in separating objects, locating positions and recovering shapes. Possible reasons for the reconstruction results are comparatively analyzed in view of artifacts. This study not only emphasizes the effects of artifacts in reconstructing multiple NLoS objects, but also predicts a promising future for our method to reconstruct the actual complex NLoS scene. In addition, we think that idea of our method might be transferred to CT field to solve the issue of aliasing artifacts. In further investigation, we are trying to reconstruct more complicated NLoS objects and develop algorithm optimizations.

## Acknowledgments

This work was supported by the National Natural Science Foundation of China (Grant No. 61102147) .